\documentclass[twocolumn,prl,aps,showpacs]{revtex4}

\usepackage{graphicx}
\usepackage{dcolumn}
\usepackage{bm}
\usepackage{txfonts}
\usepackage{pifont}
\usepackage{amssymb}
\usepackage{placeins}

\newcommand{\beginsupplement}{%
        \setcounter{table}{0}
        \renewcommand{\thetable}{S\arabic{table}}%
        \setcounter{figure}{0}
        \renewcommand{\thefigure}{S\arabic{figure}}%
     }
		
\begin{document}
\preprint{APS/123-QED}

\title{Quantum bicriticality in the heavy-fermion metamagnet YbAgGe
}

\author{Y. Tokiwa$^1$$^{\star}$}
\author{M. Garst$^2$}
\author{P. Gegenwart$^1$}
\author{S. L. Bud'ko$^3$}
\author{P. C. Canfield$^3$}

\affiliation{$^1$I. Physikalisches Institut,
Georg-August-Universit\"{a}t G\"{o}ttingen, 37077 G\"{o}ttingen,
Germany}
\affiliation{$^2$Institute for Theoretical Physics, University of Cologne, 50937 Cologne, Germany}
\affiliation{$^3$Ames Laboratory, US DOE and Department of Physics and Astronomy, Iowa State University, Ames, Iowa 50011, USA}

\date{\today}

\begin{abstract}
Bicritical points, at which two distinct symmetry-broken phases become simultaneously unstable, are typical for spin-flop metamagnetism.
Interestingly, the heavy-fermion compound YbAgGe also possesses such a bicritical point (BCP) with a low temperature $T_{\rm BCP} \approx 0.3$ K at a magnetic field of $\mu_0 H_{\rm BCP} \approx 4.5$ T. In its vicinity, YbAgGe exhibits anomalous behavior that we attribute to the influence of a quantum bicritical point (QBCP), that is close in parameter space yet can be reached by tuning $T_{\rm BCP}$ further to zero. 
Using high-resolution measurements of the magnetocaloric effect, we demonstrate that the magnetic Gr\"uneisen parameter $\Gamma_{\rm H}$ indeed both changes sign and diverges as required for quantum criticality. 
Moreover, $\Gamma_{\rm H}$ displays a characteristic scaling behavior but only on the low-field side, $H \lesssim H_{\rm BCP}$, indicating a pronounced asymmetry with respect to the critical field. 
We speculate that the small value of $T_{\rm BCP}$ is related to the geometric frustration of the Kondo-lattice of YbAgGe. 
\end{abstract}

\maketitle


Exotic quantum states of matter may arise in magnetic systems when long-range ordering of magnetic moments is suppressed by either competing interactions or geometrical frustration. A "strange metallic state", a so-called non-Fermi liquid, is found in magnetic metals close to a quantum critical point with competing Ruderman-Kittel-Kasuya-Yosida (RKKY) and Kondo interactions~\cite{LohneysenHilbertV:Ferimq}. Local-moment systems with geometrical frustration, on the other hand, remain disordered down to zero temperature and form a strongly correlated paramagnet, i.e., a quantum spin-liquid~\cite{balents-nature10}. Usually, these two intriguing quantum states of matter are experimentally studied in different material classes; the former in heavy fermions and the latter in insulating spin systems. Current interest of research on heavy fermions, however, increasingly focuses on the interplay of these two phenomena~\cite{Coleman:2010,Si:2013}, bridging two distinct fields of research. 
It has been proposed that frustrating the magnetic interaction in Kondo-lattice systems gives rise to a rich phase diagram that, as an exciting possibility, also contains unconventional, metallic spin-liquid phases~\cite{Burdin:2002,Senthil:2003}. 

For the exploration of magnetic frustration, heavy-fermion systems with a geometrically frustrated crystal lattice are of particular interest. 
The pyrochlore Kondo-lattice compound Pr$_2$Ir$_2$O$_7$, for example, indeed does not order magnetically down to lowest temperatures but exhibits anomalous thermodynamics \cite{Nakatsuji:2006} and, interestingly, a spontaneous Hall effect \cite{Machida:2009}, which have been attributed to the formation of a metallic chiral spin-liquid.

In this work, we investigate the heavy-fermion compound YbAgGe \cite{Morosan-JMMM04,budko-prb04,katoh-jmmm04,umeo-jpsj04} that possesses a hexagonal ZrNiAl-type crystal structure, forming a two-dimensional distorted Kagom\'e lattice \cite{pottgen-Kri97}, which promotes frustration effects.
At high temperatures, the magnetic susceptibility shows Curie-Weiss behavior 
with a Weiss temperature $\Theta_{ab} = -15$ K for a small magnetic field within the $ab$-plane~\cite{Morosan-JMMM04}. 
At low temperatures $T \lesssim$ 4 K, 
antiferromagnetic correlations develop, reflected in a maximum in the susceptibility $\chi_{ab}$. Finally, a sharp first-order transition into a magnetically ordered phase occurs at $T_N = 0.65$ K~\cite{Morosan-JMMM04,budko-prb04,Schmiedeshoff-prb11,tokiwa-prb06,umeo-jpsj04}.
The large factor $f = |\Theta_{ab}|/T_N = 23$ indicates frustrated interaction between magnetic moments.
This is corroborated by the structure factor of the damped spin-fluctuations observed at low temperatures, 
which consists of sheets in reciprocal space corresponding to quasi one-dimensional behavior along the $c$-axis \cite{fak-jphys05}.

\begin{figure}[]
\includegraphics[width=\linewidth,keepaspectratio]{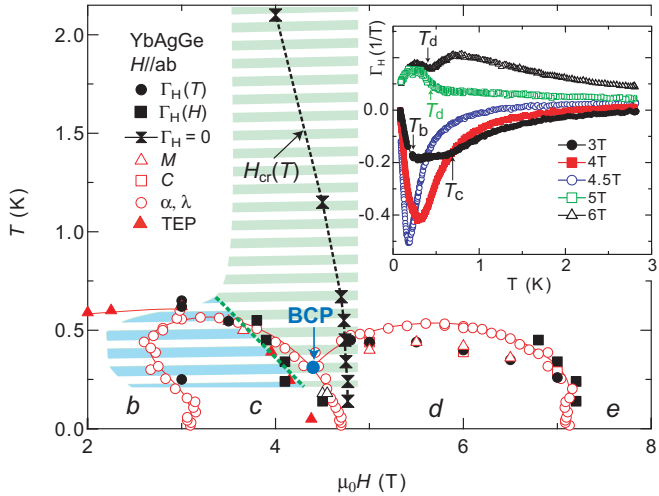}
\caption{
(Color online) $H$-$T$ phase diagram of YbAgGe for fields within the $ab$-plane focusing on the region with the bicritical point (BCP). Black and red symbols are obtained in this and previous studies~\cite{Schmiedeshoff-prb11,tokiwa-prb06,Mun-prb10}, respectively. The red solid lines are guides for the eye. The black dotted line $H_{cr}(T)$ shows maxima in entropy identified by the location of zeros of $\Gamma_{\rm H}$ 
($\blacklozenge$). 
Black solid symbols denote anomalies in $T$-sweeps (\large$\bullet$\small) and $H$-sweeps ($\blacksquare$) of $\Gamma_{\rm H}$. Red symbols are transitions observed in magnetization ($M$, $\triangle$), specific heat ($C$, $\Box$), thermal expansion coefficient and magnetostriction ($\alpha$ and $\lambda$, respectively, $\medcirc$) and thermoelectric power (TEP, $\blacktriangle$). 
Within the shaded area, $\Gamma_{\rm H}$ exhibits scaling behavior (see Fig.~\ref{fig:scalingGamma}) with a crossover indicated by the green dotted line.
Inset shows the magnetic Gr\"uneisen ratio $\Gamma_{\rm H}$ as a function of temperature for fields applied parallel to the $ab$-plane; arrows indicate phase transitions.
}
\label{fig:phasediagram}
\end{figure}

As a function of magnetic field applied within the $ab$-plane, a complex phase diagram is found with various regions 
that were labeled by letters $a$ to $f$ in Ref.~\cite{Schmiedeshoff-prb11}. The regions $a$ to $d$ are interpreted as symmetry-broken phases because their boundaries have been located by anomalies in thermodynamics and transport \cite{tokiwa-prb06,Mun-prb10,Schmiedeshoff-prb11} albeit some of them are only weak. 
The antiferromagnetic ordering wave-vectors of the $a, b$ and $c$-phases were identified by neutron scattering \cite{fak-jphys05,fak-physicab06,McMorrow-08}, whereas the order parameter of the $d$-phase remains elusive. 
It is however the latter $d$-phase that exhibits intriguing properties, see Fig.~\ref{fig:phasediagram}. 
The linear-$T$ resistivity \cite{Niklowitz-prb06}, persisting in the whole field-interval of the $d$-phase, indicates that it is a rare example of a genuine non-Fermi liquid phase. 
Furthermore, its muffin-shaped phase boundary 
implies a higher entropy than the one of the adjoining phases similar to the nematic phase in the metamagnet Sr$_3$Ru$_2$O$_7$ \cite{gegenwart-prl06,Rost-Science09}.
The anomalous behavior is most pronounced close to its lower boundary field at $\mu_0 H_{cd} = 4.8$ T, which was suggested to be caused by field-induced quantum criticality \cite{budko-prb04}. Recently, this was confirmed by the observation of a sign change in the thermal expansion and an incipient divergence of the Gr\"uneisen parameter \cite{Schmiedeshoff-prb11}. 
Furthermore, at this critical field an anomalous reduction of the Lorenz ratio was observed suggesting a violation of the Wiedemann-Franz law~\cite{Dong-prl13}. 

In Ref.~\cite{Schmiedeshoff-prb11} the origin of quantum criticality was attributed to a quantum critical endpoint (QCEP), i.e.~an isolated endpoint of a line of first-order metamagnetic quantum phase transitions \cite{millis:prl-02,Weickert-prb10,Zacharias:2013}. However, this scenario does not account for the presence of both symmetry broken phases, $c$ and $d$, close to the critical field $H_{cd}$. 
Interestingly, closer inspection of the phase diagram reveals that the phase boundaries of the $c$ and $d$-phases merge at a bicritical point (BCP) located 
at a temperature $T_{\rm BCP} \approx 0.3$ K and field $\mu_0 H_{\rm BCP} \approx 4.5$ T, see Fig.~\ref{fig:phasediagram}. Below $T_{\rm BCP}$ a direct first-order transition between the two phase emerges signaled by the appearance of hysteresis in magnetostriction \cite{Schmiedeshoff-prb11}.
As the hysteresis becomes very weak close to the BCP and the slope of the $d$-phase boundary is relatively steep so that its signatures are hard to follow in the thermal expansion \cite{Schmiedeshoff-prb11}, we estimated the location of the BCP with an uncertainty in $T_{\rm BCP}$ of about $\pm 0.1$ K.

\begin{figure}[]
\includegraphics[width=\linewidth,keepaspectratio]{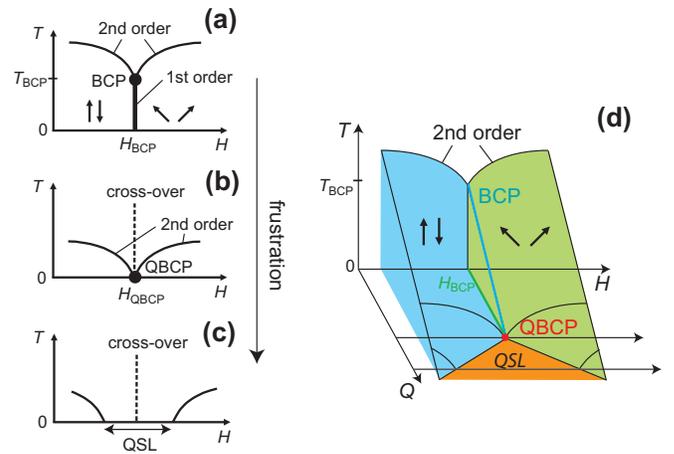}
\caption{
Schematic phase diagrams with spin-flop bicriticality at different frustration strengths $Q$. 
(a) Bicritical point at a finite temperature $T_{\rm BCP}>0$. (b) Frustration suppresses the transition temperatures and thus $T_{\rm BCP}$ leading to a quantum bicritical point (QBCP); dotted line represents a crossover line separating two paramagnetic states with different short-range order of frustrated moments. (c) Suppressing the transition temperature further results in a field-induced quantum spin-liquid (QSL) present in a finite window of magnetic field. (d) Schematic $T$-$H$-$Q$ phase diagram. Field-induced phase transition at $H_{\rm BCP}$ is of first order. QSL phase for strong frustration $Q$ is indicated by orange color.
}
\label{fig:illust}
\end{figure}

Bicriticality typically arises in local-moment antiferromagnets with weak magnetic anisotropies resulting in spin-flop metamagnetism \cite{Fisher:1974}, as illustrated in Fig.~\ref{fig:illust}(a). 
We envision that $T_{\rm BCP}$ can be tuned to zero by increasing the strength of frustration giving rise to a quantum bicritical point (QBCP), see Fig.~\ref{fig:illust}(b). 
Further increase of the frustration strength would naturally 
cause a separation of the two ordered phases leading to a 
field-induced quantum spin-liquid state existing in a finite interval of magnetic field, see Fig.~\ref{fig:illust}(c). 
We propose that in YbAgGe such a QBCP is close in parameter space and at the origin of the pronounced anomalies observed near $H_{cd}$, see Fig.~\ref{fig:illust}(d).

In order to further investigate the field-induced quantum bicriticality, we measured the magnetocaloric effect $(dT/dH)_S = T \Gamma_{\rm H}$. The quantity $\Gamma_{\rm H}$ is a magnetic analogue of the Gr\"uneisen parameter, that necessarily diverges at a quantum critical point (QCP) \cite{zhu,GarstM:SigctG}. Compared to the latter, that was already determined for YbAgGe in Ref.~\cite{Schmiedeshoff-prb11}, $\Gamma_{\rm H}$ has the important advantage that it does not necessitate additional assumptions about the magnetoelastic coupling, and it directly provides information about the entropy distribution within the $H$-$T$ phase diagram via the relation $(\partial S/\partial H)_T = - \Gamma_{\rm H}/C_H$. In particular, maxima in entropy are reflected in sign changes of $\Gamma_{\rm H}$ \cite{GarstM:SigctG}. 
For the prototypical field-induced QCP material YbRh$_2$Si$_2$, a divergence of $\Gamma_{\rm H}$ at its critical field was reported in Ref.~\cite{tokiwa-prl09}.  

Single crystals were grown from high temperature ternary solutions rich in Ag and Ge~\cite{Morosan-JMMM04}. The magnetocaloric effect was measured with very high resolution in a dilution refrigerator with a SC magnet equipped with an additional modulation coil by utilizing the alternating field technique~\cite{tokiwa-rsi11}. 
 

Temperature scans of $\Gamma_{\rm H}$ of YbAgGe are shown in the inset of Fig.~\ref{fig:phasediagram} at different magnetic fields 3\,T $\leq\mu_0 H\leq$ 6\,T. The arrows indicate the anomalies associated with the 
phase transitions into the $c$ and $d$-phases. 
The most noticeable feature, however, is the spreading of the set of $\Gamma_{\rm H}$ curves as the temperature is lowered down to the transition temperatures with a sign change around $H_{cd}$. Such a behavior originates from an accumulation of entropy close to the critical magnetic field and is characteristic for metamagnetic quantum criticality \cite{millis:prl-02,Weickert-prb10,Zacharias:2013}.
At the low-field side $H < H_{cd}$ for $\mu_0 H = 4$ and 4.5 T, $\Gamma_{\rm H}$ is strongly temperature dependent
and develops a pronounced peak whose height and width increases and decreases, respectively, as criticality is approached. The peak position for $\mu_0 H = 4.5$ T is located at 0.18\,K well below the transition temperature 
indicating that the quantum critical fluctuations are hardly quenched upon entering the ordered phase. In contrast, for $H > H_{cd}$ such a peak is absent as $\Gamma_{\rm H}(T)$ is substantially suppressed upon entering the $d$-phase, e.g., at $6$ T. 

\begin{figure}[t]
\includegraphics[width=\linewidth,keepaspectratio]{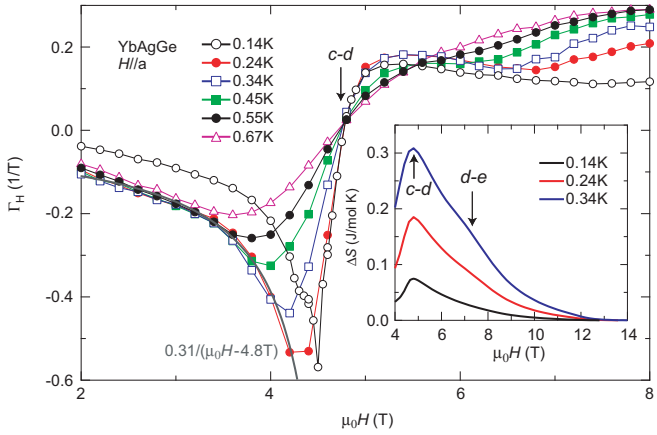}
\caption{
(Color online) $\Gamma_{\rm H}$ as a function of magnetic field applied parallel to the 
$ab$-plane for low temperatures $T < 0.7$~K. Grey solid line is the power law 0.31/($\mu_0H$-4.8 T). Inset shows change of entropy calculated from $\Gamma_{\rm H}$ with the help of the reported specific heat~\cite{tokiwa-prb06} (normalized such that $\Delta S = 0$ for $\mu_0 H = 13$ T). Arrows indicate the positions of phase transitions at $T=0$.
}
\label{fig:GammaH-Hsweep}
\end{figure}

Magnetic field-sweeps of $\Gamma_{\rm H}$ for temperatures $T < 0.7$ K are shown in Fig.~\ref{fig:GammaH-Hsweep}. Whereas the signatures at the transitions $b$-$c$ and $d$-$e$ are minor, the transition between $d$ and $c$ is clearly revealed. In the field-sweep, the sign change of $\Gamma_{\rm H}$ close to the critical field $H_{cd}$ is particularly evident. A sign change of $\Gamma_{\rm H}$ coincides with a maximum in entropy $S(H)$ as illustrated in the inset of Fig.~\ref{fig:GammaH-Hsweep}. Here, the entropy was obtained by integrating $(\partial S/\partial H)_T = - \Gamma_{\rm H}/C_H$ using the specific heat data of Ref.~\cite{tokiwa-prb06}.
The position of sign changes, $\Gamma_{\rm H}=0$, obtained from the $T$- and $H$-sweeps defines 
the entropy ridge $H_{cr}(T)$ that is shown by the black dotted line in Fig.~\ref{fig:phasediagram}. In the low temperature limit it extrapolates to the critical field, $H_{cr}(T) \to H_{cd}$ for $T\to 0$. Its temperature dependence, $H_{cr}(T)$, is rather weak as it starts with an infinite slope. Nevertheless, it is considerably stronger (20\% reduction at 2K from its zero temperature limit) as compared to other itinerant quantum critical metamagnets like CeRu$_2$Si$_2$ ($\sim$0.5\% increase at 1.5K)~\cite{Weickert-prb10} and Sr$_3$Ru$_2$O$_7$ (no change up to 3K)~\cite{gegenwart-prl06}.
 
As already anticipated from the $T$-sweeps, the field dependence of $\Gamma_{\rm H}$ around the critical field becomes highly asymmetric at low temperatures with 
a rounded shoulder for $H > H_{cd}$ but a pronounced negative peak on the low-field side. Strikingly, for temperatures in the range 0.24 up to 0.64 K $\Gamma_{\rm H}(H)$ first nicely traces a common curve $\Gamma_{\rm H} \approx -G_{\rm H}/(H-H_{cd})$ as expected for quantum criticality \cite{zhu} with a fitted prefactor of $G_{\rm H} \approx -0.31$. Closer to the critical field it crosses over towards positive values with a sign change at the critical field that persists down to lowest temperatures. 
Finally at a temperature T = 0.14 K well below the bicritical point $\Gamma_{\rm H}$ still exhibits a sharp decrease towards the critical field but deviates from the common scaling curve.

 \begin{figure}[t]
\includegraphics[width=\linewidth,keepaspectratio]{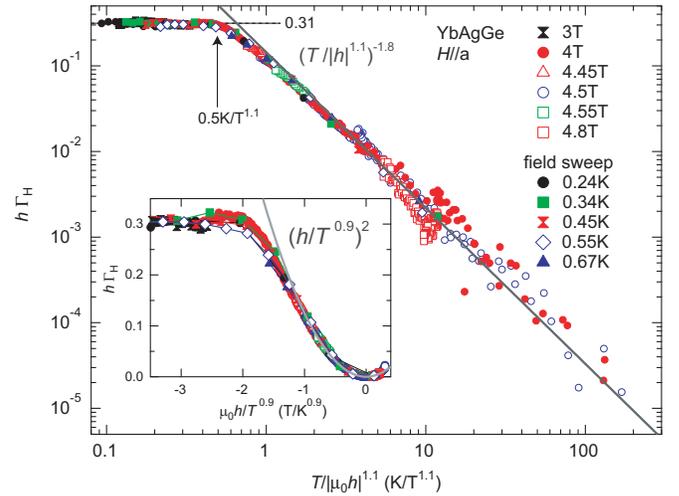}
\caption{
(Color online) Scaling plot of $h \Gamma_{\rm H}$ vs $T/|h|^{1.1}$ for data only within the shaded regime of Fig.~\ref{fig:phasediagram} with $h=H-H_{cr}(T)$ where $H_{cr}(T)$ is defined by the zeros of $\Gamma_{\rm H}$ (dotted line in Fig.~\ref{fig:phasediagram}). An arrow indicates the crossover at $0.5$\,K/T$^{1.1}$ of the scaling function also shown by the green dotted line in Fig.~\ref{fig:phasediagram}.
Grey solid line represents the asymptotics $\Gamma_{\rm H} \sim h/T^{1.8}$. Inset shows the scaling plot on a linear scale but as a function of $h/T^{0.9}$. 
}
\label{fig:scalingGamma}
\end{figure}

In the case of a metamagnetic quantum critical endpoint (QCEP)
an Ising symmetry emerges resulting in symmetric behavior between low- and high fields \cite{Weickert-prb10}. Such an approximate symmetry is observed in Sr$_3$Ru$_2$O$_7$ \cite{gegenwart-prl06}, CeRu$_2$Si$_2$ \cite{Weickert-prb10}, and Ca$_{2-x}$Sr$_x$RuO$_4$ \cite{Baier:2007}. 
The QBCP also terminates a line of first-order transitions that, in contrast to the QCEP, 
separates however two distinct symmetry-broken phases. As a consequence, quantum bicritical behavior is generically expected to be asymmetric with respect to the critical field, which apparently applies to YbAgGe. Such an asymmetry might be induced,  in particular, by distinct dynamical exponents, $z$, for critical fluctuations associated with the two adjacent phases, so that the QBCP is generally characterized by multiple dynamics \cite{Zacharias:2009,Meng:2012}.  
 
In order to investigate the properties of the QBCP in YbAgGe quantitatively, we proceed with analyzing the scaling of $\Gamma_{\rm H}$ observed experimentally.  As the critical signatures for high fields are rather weak we concentrate in the following on the scaling that is apparent on the low-field side, $H<H_{cd}$. Similarly as suggested for the QCEP \cite{Zacharias:2013}, we consider two scaling parameters given by temperature $T$ and $h = H-H_{cr}(T)$, i.e. the distance in field to the location of entropy maxima, $H_{cr}(T)$. The weak $T$-dependence of $H_{cr}(T)$ is 
irrelevant in the limit $T \to 0$ but the extracted scaling sensitively depends on it. 
Using the temperature dependent scaling field $h$, it is possible to reveal scaling behavior in a larger temperature regime otherwise hidden by the temperature drift of the entropy ridge $H_{cr}(T)$.

For data on the low-field side, $H < H_{cd}$, and $T> 0.2$\,K within the shaded regime of Fig.~\ref{fig:phasediagram}, we find that $\Gamma_{\rm H}$ obeys $T/|h|^{1.1}$ scaling behavior, see Fig.~\ref{fig:scalingGamma}. It can be described by a function of the form $\Gamma_{\rm H} = \frac{1}{h} \mathcal{G}(h/T^{1/(\nu z)})$ with an exponent $\nu z = 1.1$. The excellent collapse of data with the absolute temperature $T$ as a scaling field confirms in particular that 
the anomalous behavior around $H_{cd}$ is indeed caused by a {\it quantum} (bi-)critical point and not by the classical counterpart at $T_{\rm BCP} \approx 0.3$\,K. For classical bicriticality instead a scaling collapse in terms 
of the differences $T-T_{\rm BCP}$ and $H-H_{\rm BCP}$ would be expected. 

Two scaling regimes separated by a crossover at $T/|\mu_0h|^{1.1} \approx 0.5$\,K/T$^{1.1}$ can be distinguished. For low temperatures and large negative $h$, $\Gamma_{\rm H} h$ approaches a constant 
with the value $-G_{\rm H} = 0.31$. For high temperatures and small $h$, on the other hand, it is expected that $\Gamma_{\rm H}$ vanishes analytically, i.e., linearly with $h$ because the line $H_{cr}(T)$ is only a crossover where thermodynamics remains smooth \cite{markus_note}. 
Analyticity thus requires the function $\mathcal{G}$ to behave for small arguments as $\mathcal{G}(x) \propto x^2$. This determines the asymptotics $\Gamma_{\rm H} \sim h/T^{2/(\nu z)} \approx h/T^{1.8}$, which is indeed observed at high temperatures as indicated by the solid line in Fig.~\ref{fig:scalingGamma}.

Interestingly, in the mixed-valence compound $\beta$-YbAlB$_4$ a $T/h$ scaling with a similar exponent ($\nu z$=1) and a divergence $\Gamma_H \sim h/T^2$ was observed but with $H_{cr} = 0$ \cite{Matsumoto:2011}. Effective one-dimensional degrees of freedom have been invoked there for its explanation \cite{Ramires:2012}. 
Quasi one-dimensional fluctuations have been already observed in YbAgGe but in zero field probably promoted by the geometrical frustration \cite{fak-jphys05}. This low dimensionality of fluctuations might also survive in finite field and drive the critical behavior at $H_{cd}$. A hint in this direction is provided by the strong temperature dependence of the magnetization $(\partial M/\partial T)_H$~\cite{tokiwa-prb06} and the thermal expansion $\alpha$~\cite{Schmiedeshoff-prb11} that both behave as $\sim h/T^{1.7}$ close to the critical field \cite{SM}. From a quantum critical scaling point of view such a strong divergence as a function of $T$ implies a low spatial dimensionality $d$~\cite{zhu}.

A neutron scattering study in the bicritical regime would be useful not only to determine the dimensionality but also the dynamics of the critical fluctuations. This might also shed light on the pronounced asymmetry of the quantum bicritical behavior with respect to the critical field. Moreover, in order to construct a basic theory of the QBCP and to test its spin-flop character, the experimental identification of the order parameter of the $d$-phase is mandatory. Further experimental and theoretical work will be required to elucidate the origin of the QBCP and the observed scaling exponents.

To summarize, we propose that YbAgGe is situated close to a quantum bicritical point (QBCP) that controls thermodynamics in its vicinity. 
We verified that the magnetic Gr\"uneisen parameter $\Gamma_{\rm H}$ exhibits the corresponding quantum critical signatures and identified a characteristic scaling behavior.

We would like to acknowledge helpful discussions with G. M. Schmiedeshoff. This work has been supported by the German Science Foundation
through FOR 960 (Quantum phase transitions). Part of this research was performed by P.~C.~Canfield and S.~L.~Bud'ko at the Ames Laboratory and supported by the U.S. Department of Energy, Office of Basic Energy Science, Division of Materials
Sciences and Engineering. Ames Laboratory is operated for the U.S. Department of Energy by Iowa State University under Contract No. DE-AC02-07CH11358.

\beginsupplement

\section{Supplementary Material for: Quantum bicriticality in the heavy-fermion metamagnet ${\bf YbAgGe}$}

\begin{figure}[ht*]
\includegraphics[width=\linewidth,keepaspectratio]{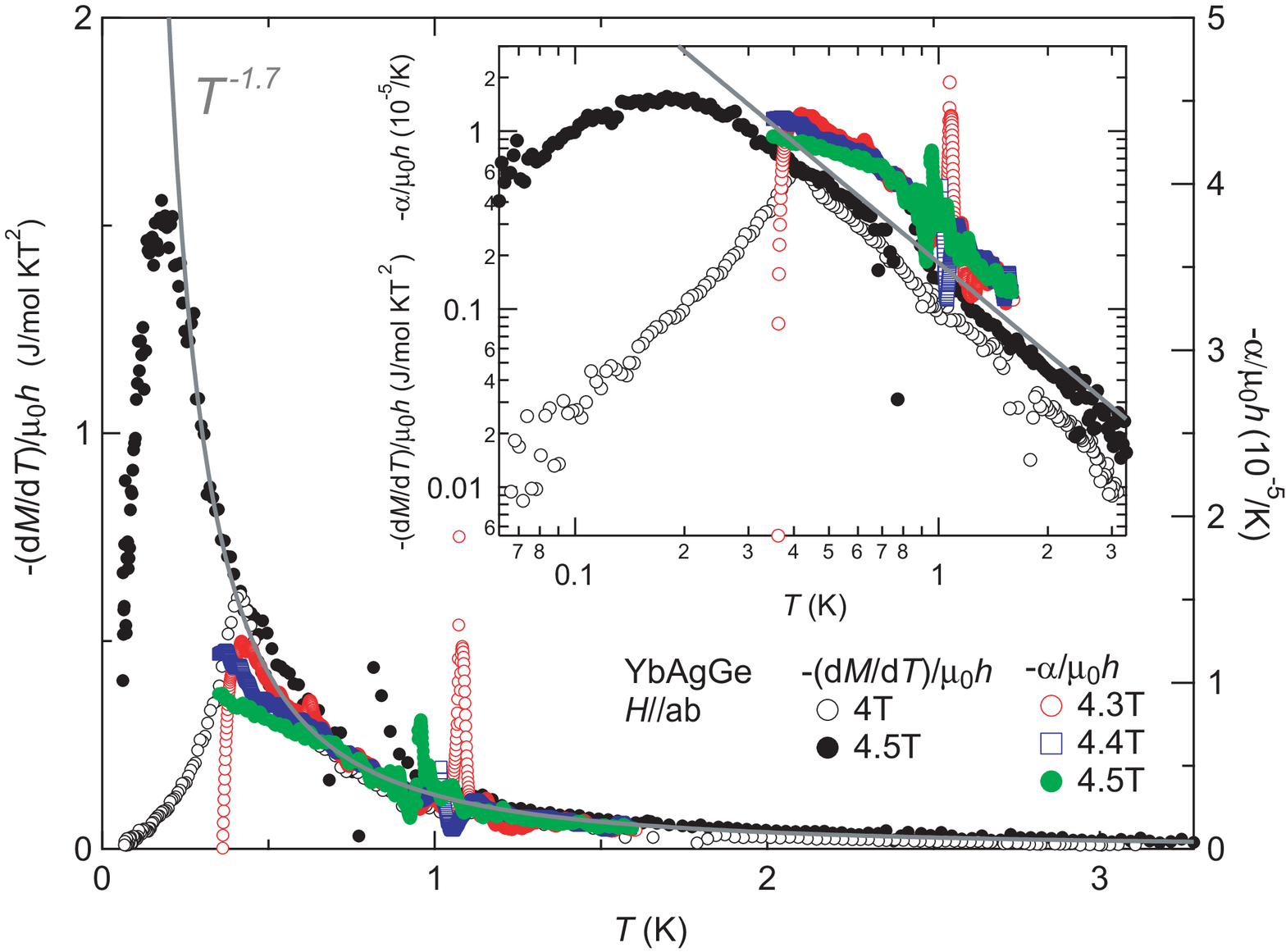}
\caption{(Color online) Plot of the temperature derivative of the magnetization $(\partial M/\partial T)_H$~\cite{tokiwa-prb06} and the thermal expansion $\alpha$~\cite{Schmiedeshoff-prb11} divided by $h= H-H_{cr}(T)$ vs. temperature $T$. Spiky features (for example at around 1.1\,K seen in $\alpha /\mu_0h$ for $H$=4.3\,T) are artifacts, caused by vanishing $h$, when it approaches $H_{cr}(T)$ line in the phase diagram (see Fig.~\ref{fig:phasediagram}). Both quantities diverge strongly with decreasing $T$, $(\partial M/\partial T)_H \sim \alpha \sim h/T^{1.7}$. Inset shows the same data on a double logarithmic plot.
}
\label{fig:scalingAlpha}
\end{figure}

We provide an analysis of data for the temperature derivative of the magnetization $(\partial M/\partial T)_H$ and the thermal expansion $\alpha$ already published in Refs.~\cite{tokiwa-prb06} and \cite{Schmiedeshoff-prb11}, respectively. Both quantities are expected to exhibit the same metamagnetic singularities. As both quantities vanish at the locations of maximum entropy, $H_{cr}(T)$, we have divided them by $h = H-H_{\rm cr}(T)$ resulting in the spiky artifacts at the temperatures where $h=0$, see Fig.~\ref{fig:scalingAlpha}. Displaying the data in this way, however, reveals strikingly that both quantities obey the power-law $(\partial M/\partial T)_H \sim \alpha \sim h/T^{1.7}$ and thus increase strongly with decreasing temperatures.

\begin{thebibliography}{38}
\expandafter\ifx\csname natexlab\endcsname\relax\def\natexlab#1{#1}\fi
\expandafter\ifx\csname bibnamefont\endcsname\relax
  \def\bibnamefont#1{#1}\fi
\expandafter\ifx\csname bibfnamefont\endcsname\relax
  \def\bibfnamefont#1{#1}\fi
\expandafter\ifx\csname citenamefont\endcsname\relax
  \def\citenamefont#1{#1}\fi
\expandafter\ifx\csname url\endcsname\relax
  \def\url#1{\texttt{#1}}\fi
\expandafter\ifx\csname urlprefix\endcsname\relax\def\urlprefix{URL }\fi
\providecommand{\bibinfo}[2]{#2}
\providecommand{\eprint}[2][]{\url{#2}}

\bibitem[{\citenamefont{v.~L\"ohneysen
  et~al.}(2007)\citenamefont{v.~L\"ohneysen, Rosch, Vojta, and
  W\"olfle}}]{LohneysenHilbertV:Ferimq}
\bibinfo{author}{\bibfnamefont{H.}~\bibnamefont{v.~L\"ohneysen}},
  \bibinfo{author}{\bibfnamefont{A.}~\bibnamefont{Rosch}},
  \bibinfo{author}{\bibfnamefont{M.}~\bibnamefont{Vojta}}, \bibnamefont{and}
  \bibinfo{author}{\bibfnamefont{P.}~\bibnamefont{W\"olfle}},
  \bibinfo{journal}{Rev. Mod. Phys.} \textbf{\bibinfo{volume}{79}},
  \bibinfo{pages}{1015 } (\bibinfo{year}{2007}).

\bibitem[{\citenamefont{Balents}(2010)}]{balents-nature10}
\bibinfo{author}{\bibfnamefont{L.}~\bibnamefont{Balents}},
  \bibinfo{journal}{Nature} \textbf{\bibinfo{volume}{464}},
  \bibinfo{pages}{199} (\bibinfo{year}{2010}).

\bibitem[{\citenamefont{Coleman and Nevidomskyy}(2010)}]{Coleman:2010}
\bibinfo{author}{\bibfnamefont{P.}~\bibnamefont{Coleman}} \bibnamefont{and}
  \bibinfo{author}{\bibfnamefont{A.~H.} \bibnamefont{Nevidomskyy}},
  \bibinfo{journal}{J. Low Temp. Phys.} \textbf{\bibinfo{volume}{161}},
  \bibinfo{pages}{182} (\bibinfo{year}{2010}).

\bibitem[{\citenamefont{Si and Paschen}(2013)}]{Si:2013}
\bibinfo{author}{\bibfnamefont{Q.}~\bibnamefont{Si}} \bibnamefont{and}
  \bibinfo{author}{\bibfnamefont{S.}~\bibnamefont{Paschen}},
  \bibinfo{journal}{Physica Status Solidi (b)} \textbf{\bibinfo{volume}{250}},
  \bibinfo{pages}{425} (\bibinfo{year}{2013}).

\bibitem[{\citenamefont{Burdin et~al.}(2002)\citenamefont{Burdin, Grempel, and
  Georges}}]{Burdin:2002}
\bibinfo{author}{\bibfnamefont{S.}~\bibnamefont{Burdin}},
  \bibinfo{author}{\bibfnamefont{D.}~\bibnamefont{Grempel}}, \bibnamefont{and}
  \bibinfo{author}{\bibfnamefont{A.}~\bibnamefont{Georges}},
  \bibinfo{journal}{Phys. Rev. B} \textbf{\bibinfo{volume}{66}},
  \bibinfo{pages}{045111} (\bibinfo{year}{2002}).

\bibitem[{\citenamefont{Senthil et~al.}(2003)\citenamefont{Senthil, Sachdev,
  and Vojta}}]{Senthil:2003}
\bibinfo{author}{\bibfnamefont{T.}~\bibnamefont{Senthil}},
  \bibinfo{author}{\bibfnamefont{S.}~\bibnamefont{Sachdev}}, \bibnamefont{and}
  \bibinfo{author}{\bibfnamefont{M.}~\bibnamefont{Vojta}},
  \bibinfo{journal}{Phys. Rev. Lett.} \textbf{\bibinfo{volume}{90}},
  \bibinfo{pages}{216403} (\bibinfo{year}{2003}).

\bibitem[{\citenamefont{Nakatsuji et~al.}(2006)\citenamefont{Nakatsuji,
  Machida, Maeno, Tayama, Sakakibara, van Duijn, Balicas, Millican, Macaluso,
  and Chan}}]{Nakatsuji:2006}
\bibinfo{author}{\bibfnamefont{S.}~\bibnamefont{Nakatsuji}},
  \bibinfo{author}{\bibfnamefont{Y.}~\bibnamefont{Machida}},
  \bibinfo{author}{\bibfnamefont{Y.}~\bibnamefont{Maeno}},
  \bibinfo{author}{\bibfnamefont{T.}~\bibnamefont{Tayama}},
  \bibinfo{author}{\bibfnamefont{T.}~\bibnamefont{Sakakibara}},
  \bibinfo{author}{\bibfnamefont{J.}~\bibnamefont{van Duijn}},
  \bibinfo{author}{\bibfnamefont{L.}~\bibnamefont{Balicas}},
  \bibinfo{author}{\bibfnamefont{J.~N.} \bibnamefont{Millican}},
  \bibinfo{author}{\bibfnamefont{R.~T.} \bibnamefont{Macaluso}},
  \bibnamefont{and} \bibinfo{author}{\bibfnamefont{J.~Y.} \bibnamefont{Chan}},
  \bibinfo{journal}{Phys. Rev. Lett.} \textbf{\bibinfo{volume}{96}},
  \bibinfo{pages}{87204} (\bibinfo{year}{2006}).

\bibitem[{\citenamefont{Machida et~al.}(2009)\citenamefont{Machida, Nakatsuji,
  Onoda, Tayama, and Sakakibara}}]{Machida:2009}
\bibinfo{author}{\bibfnamefont{Y.}~\bibnamefont{Machida}},
  \bibinfo{author}{\bibfnamefont{S.}~\bibnamefont{Nakatsuji}},
  \bibinfo{author}{\bibfnamefont{S.}~\bibnamefont{Onoda}},
  \bibinfo{author}{\bibfnamefont{T.}~\bibnamefont{Tayama}}, \bibnamefont{and}
  \bibinfo{author}{\bibfnamefont{T.}~\bibnamefont{Sakakibara}},
  \bibinfo{journal}{Nature} \textbf{\bibinfo{volume}{463}},
  \bibinfo{pages}{210} (\bibinfo{year}{2009}).

\bibitem[{\citenamefont{Morosan et~al.}(2004)\citenamefont{Morosan, Bud'ko,
  Canfield, Torikachvili, and Lacerda}}]{Morosan-JMMM04}
\bibinfo{author}{\bibfnamefont{E.}~\bibnamefont{Morosan}},
  \bibinfo{author}{\bibfnamefont{S.}~\bibnamefont{Bud'ko}},
  \bibinfo{author}{\bibfnamefont{P.}~\bibnamefont{Canfield}},
  \bibinfo{author}{\bibfnamefont{M.}~\bibnamefont{Torikachvili}},
  \bibnamefont{and} \bibinfo{author}{\bibfnamefont{A.}~\bibnamefont{Lacerda}},
  \bibinfo{journal}{J. Magn. Magn. Mater.} \textbf{\bibinfo{volume}{277}},
  \bibinfo{pages}{298 } (\bibinfo{year}{2004}).

\bibitem[{\citenamefont{Bud'ko et~al.}(2004)\citenamefont{Bud'ko, Morosan, and
  Canfield}}]{budko-prb04}
\bibinfo{author}{\bibfnamefont{S.~L.} \bibnamefont{Bud'ko}},
  \bibinfo{author}{\bibfnamefont{E.}~\bibnamefont{Morosan}}, \bibnamefont{and}
  \bibinfo{author}{\bibfnamefont{P.~C.} \bibnamefont{Canfield}},
  \bibinfo{journal}{Phys. Rev. B} \textbf{\bibinfo{volume}{69}},
  \bibinfo{pages}{014415} (\bibinfo{year}{2004}).

\bibitem[{\citenamefont{Katoh et~al.}(2004)\citenamefont{Katoh, Mano, Nakano,
  Terui, Niide, and Ochiai}}]{katoh-jmmm04}
\bibinfo{author}{\bibfnamefont{K.}~\bibnamefont{Katoh}},
  \bibinfo{author}{\bibfnamefont{Y.}~\bibnamefont{Mano}},
  \bibinfo{author}{\bibfnamefont{K.}~\bibnamefont{Nakano}},
  \bibinfo{author}{\bibfnamefont{G.}~\bibnamefont{Terui}},
  \bibinfo{author}{\bibfnamefont{Y.}~\bibnamefont{Niide}}, \bibnamefont{and}
  \bibinfo{author}{\bibfnamefont{A.}~\bibnamefont{Ochiai}},
  \bibinfo{journal}{J. Magn. Magn. Mater.} \textbf{\bibinfo{volume}{268}},
  \bibinfo{pages}{212 } (\bibinfo{year}{2004}).

\bibitem[{\citenamefont{Umeo et~al.}(2004)\citenamefont{Umeo, Yamane, Muro,
  Katoh, Niide, Ochiai, Morie, Sakakibara, and Takabatake}}]{umeo-jpsj04}
\bibinfo{author}{\bibfnamefont{K.}~\bibnamefont{Umeo}},
  \bibinfo{author}{\bibfnamefont{K.}~\bibnamefont{Yamane}},
  \bibinfo{author}{\bibfnamefont{Y.}~\bibnamefont{Muro}},
  \bibinfo{author}{\bibfnamefont{K.}~\bibnamefont{Katoh}},
  \bibinfo{author}{\bibfnamefont{Y.}~\bibnamefont{Niide}},
  \bibinfo{author}{\bibfnamefont{A.}~\bibnamefont{Ochiai}},
  \bibinfo{author}{\bibfnamefont{T.}~\bibnamefont{Morie}},
  \bibinfo{author}{\bibfnamefont{T.}~\bibnamefont{Sakakibara}},
  \bibnamefont{and}
  \bibinfo{author}{\bibfnamefont{T.}~\bibnamefont{Takabatake}},
  \bibinfo{journal}{J. Phys. Soc. Jpn.} \textbf{\bibinfo{volume}{73}},
  \bibinfo{pages}{537} (\bibinfo{year}{2004}).

\bibitem[{\citenamefont{Pottgen et~al.}(1997)\citenamefont{Pottgen, Gibson, and
  Kremer}}]{pottgen-Kri97}
\bibinfo{author}{\bibfnamefont{R.}~\bibnamefont{Pottgen}},
  \bibinfo{author}{\bibfnamefont{B.}~\bibnamefont{Gibson}}, \bibnamefont{and}
  \bibinfo{author}{\bibfnamefont{R.~K.} \bibnamefont{Kremer}},
  \bibinfo{journal}{Z. Kristallogr. - New Cryst. Struct.}
  \textbf{\bibinfo{volume}{212}}, \bibinfo{pages}{58} (\bibinfo{year}{1997}).

\bibitem[{\citenamefont{Schmiedeshoff et~al.}(2011)\citenamefont{Schmiedeshoff,
  Mun, Lounsbury, Tracy, Palm, Hannahs, Park, Murphy, Bud'ko, and
  Canfield}}]{Schmiedeshoff-prb11}
\bibinfo{author}{\bibfnamefont{G.~M.} \bibnamefont{Schmiedeshoff}},
  \bibinfo{author}{\bibfnamefont{E.~D.} \bibnamefont{Mun}},
  \bibinfo{author}{\bibfnamefont{A.~W.} \bibnamefont{Lounsbury}},
  \bibinfo{author}{\bibfnamefont{S.~J.} \bibnamefont{Tracy}},
  \bibinfo{author}{\bibfnamefont{E.~C.} \bibnamefont{Palm}},
  \bibinfo{author}{\bibfnamefont{S.~T.} \bibnamefont{Hannahs}},
  \bibinfo{author}{\bibfnamefont{J.-H.} \bibnamefont{Park}},
  \bibinfo{author}{\bibfnamefont{T.~P.} \bibnamefont{Murphy}},
  \bibinfo{author}{\bibfnamefont{S.~L.} \bibnamefont{Bud'ko}},
  \bibnamefont{and} \bibinfo{author}{\bibfnamefont{P.~C.}
  \bibnamefont{Canfield}}, \bibinfo{journal}{Phys. Rev. B}
  \textbf{\bibinfo{volume}{83}}, \bibinfo{pages}{180408}
  (\bibinfo{year}{2011}).

\bibitem[{\citenamefont{Tokiwa et~al.}(2006)\citenamefont{Tokiwa, Pikul,
  Gegenwart, Steglich, Bud'ko, and Canfield}}]{tokiwa-prb06}
\bibinfo{author}{\bibfnamefont{Y.}~\bibnamefont{Tokiwa}},
  \bibinfo{author}{\bibfnamefont{A.}~\bibnamefont{Pikul}},
  \bibinfo{author}{\bibfnamefont{P.}~\bibnamefont{Gegenwart}},
  \bibinfo{author}{\bibfnamefont{F.}~\bibnamefont{Steglich}},
  \bibinfo{author}{\bibfnamefont{S.~L.} \bibnamefont{Bud'ko}},
  \bibnamefont{and} \bibinfo{author}{\bibfnamefont{P.~C.}
  \bibnamefont{Canfield}}, \bibinfo{journal}{Phys. Rev. B}
  \textbf{\bibinfo{volume}{73}}, \bibinfo{pages}{094435}
  (\bibinfo{year}{2006}).

\bibitem[{\citenamefont{F\r{a}k et~al.}(2005)\citenamefont{F\r{a}k, McMorrow,
  Niklowitz, Raymond, Ressouche, Flouquet, Canfield, Bud'ko, Janssen, and
  Gutmann}}]{fak-jphys05}
\bibinfo{author}{\bibfnamefont{B.}~\bibnamefont{F\r{a}k}},
  \bibinfo{author}{\bibfnamefont{D.~F.} \bibnamefont{McMorrow}},
  \bibinfo{author}{\bibfnamefont{P.~G.} \bibnamefont{Niklowitz}},
  \bibinfo{author}{\bibfnamefont{S.}~\bibnamefont{Raymond}},
  \bibinfo{author}{\bibfnamefont{E.}~\bibnamefont{Ressouche}},
  \bibinfo{author}{\bibfnamefont{J.}~\bibnamefont{Flouquet}},
  \bibinfo{author}{\bibfnamefont{P.~C.} \bibnamefont{Canfield}},
  \bibinfo{author}{\bibfnamefont{S.~L.} \bibnamefont{Bud'ko}},
  \bibinfo{author}{\bibfnamefont{Y.}~\bibnamefont{Janssen}}, \bibnamefont{and}
  \bibinfo{author}{\bibfnamefont{M.~J.} \bibnamefont{Gutmann}},
  \bibinfo{journal}{J. Phys.: Condensed Matter} \textbf{\bibinfo{volume}{17}},
  \bibinfo{pages}{301} (\bibinfo{year}{2005}).

\bibitem[{\citenamefont{Mun et~al.}(2010)\citenamefont{Mun, Bud'ko, and
  Canfield}}]{Mun-prb10}
\bibinfo{author}{\bibfnamefont{E.}~\bibnamefont{Mun}},
  \bibinfo{author}{\bibfnamefont{S.~L.} \bibnamefont{Bud'ko}},
  \bibnamefont{and} \bibinfo{author}{\bibfnamefont{P.~C.}
  \bibnamefont{Canfield}}, \bibinfo{journal}{Phys. Rev. B}
  \textbf{\bibinfo{volume}{82}}, \bibinfo{pages}{174403}
  (\bibinfo{year}{2010}).

\bibitem[{\citenamefont{F\r{a}k et~al.}(2006)\citenamefont{F\r{a}k, Ruegg,
  Niklowitz, McMorrow, Canfield, Bud'ko, Janssen, and
  Habicht}}]{fak-physicab06}
\bibinfo{author}{\bibfnamefont{B.}~\bibnamefont{F\r{a}k}},
  \bibinfo{author}{\bibfnamefont{C.}~\bibnamefont{Ruegg}},
  \bibinfo{author}{\bibfnamefont{P.}~\bibnamefont{Niklowitz}},
  \bibinfo{author}{\bibfnamefont{D.}~\bibnamefont{McMorrow}},
  \bibinfo{author}{\bibfnamefont{P.}~\bibnamefont{Canfield}},
  \bibinfo{author}{\bibfnamefont{S.}~\bibnamefont{Bud'ko}},
  \bibinfo{author}{\bibfnamefont{Y.}~\bibnamefont{Janssen}}, \bibnamefont{and}
  \bibinfo{author}{\bibfnamefont{K.}~\bibnamefont{Habicht}},
  \bibinfo{journal}{Physica B: Condensed Matter}
  \textbf{\bibinfo{volume}{378-380}}, \bibinfo{pages}{669 }
  (\bibinfo{year}{2006}).

\bibitem[{\citenamefont{McMorrow~{\it et al.}}(2008)}]{McMorrow-08}
\bibinfo{author}{\bibfnamefont{D.~F.} \bibnamefont{McMorrow~{\it et al.}}},
  \bibinfo{journal}{Proceedings of the 25th International Conference on Low
  Temperature Physics, Amsterdam (unpublished)}  (\bibinfo{year}{2008}).

\bibitem[{\citenamefont{Niklowitz et~al.}(2006)\citenamefont{Niklowitz, Knebel,
  Flouquet, Bud'ko, and Canfield}}]{Niklowitz-prb06}
\bibinfo{author}{\bibfnamefont{P.~G.} \bibnamefont{Niklowitz}},
  \bibinfo{author}{\bibfnamefont{G.}~\bibnamefont{Knebel}},
  \bibinfo{author}{\bibfnamefont{J.}~\bibnamefont{Flouquet}},
  \bibinfo{author}{\bibfnamefont{S.~L.} \bibnamefont{Bud'ko}},
  \bibnamefont{and} \bibinfo{author}{\bibfnamefont{P.~C.}
  \bibnamefont{Canfield}}, \bibinfo{journal}{Phys. Rev. B}
  \textbf{\bibinfo{volume}{73}}, \bibinfo{pages}{125101}
  (\bibinfo{year}{2006}).

\bibitem[{\citenamefont{Gegenwart et~al.}(2006)\citenamefont{Gegenwart,
  Weickert, Garst, Perry, and Maeno}}]{gegenwart-prl06}
\bibinfo{author}{\bibfnamefont{P.}~\bibnamefont{Gegenwart}},
  \bibinfo{author}{\bibfnamefont{F.}~\bibnamefont{Weickert}},
  \bibinfo{author}{\bibfnamefont{M.}~\bibnamefont{Garst}},
  \bibinfo{author}{\bibfnamefont{R.~S.} \bibnamefont{Perry}}, \bibnamefont{and}
  \bibinfo{author}{\bibfnamefont{Y.}~\bibnamefont{Maeno}},
  \bibinfo{journal}{Phys. Rev. Lett.} \textbf{\bibinfo{volume}{96}},
  \bibinfo{pages}{136402} (\bibinfo{year}{2006}).

\bibitem[{\citenamefont{Rost et~al.}(2009)\citenamefont{Rost, Perry, Mercure,
  Mackenzie, and Grigera}}]{Rost-Science09}
\bibinfo{author}{\bibfnamefont{A.~W.} \bibnamefont{Rost}},
  \bibinfo{author}{\bibfnamefont{R.~S.} \bibnamefont{Perry}},
  \bibinfo{author}{\bibfnamefont{J.-F.} \bibnamefont{Mercure}},
  \bibinfo{author}{\bibfnamefont{A.~P.} \bibnamefont{Mackenzie}},
  \bibnamefont{and} \bibinfo{author}{\bibfnamefont{S.~A.}
  \bibnamefont{Grigera}}, \bibinfo{journal}{Science}
  \textbf{\bibinfo{volume}{325}}, \bibinfo{pages}{1360} (\bibinfo{year}{2009}).

\bibitem[{\citenamefont{Dong et~al.}(2013)\citenamefont{Dong, Tokiwa, Bud'ko,
  Canfield, and Gegenwart}}]{Dong-prl13}
\bibinfo{author}{\bibfnamefont{J.~K.} \bibnamefont{Dong}},
  \bibinfo{author}{\bibfnamefont{Y.}~\bibnamefont{Tokiwa}},
  \bibinfo{author}{\bibfnamefont{S.~L.} \bibnamefont{Bud'ko}},
  \bibinfo{author}{\bibfnamefont{P.~C.} \bibnamefont{Canfield}},
  \bibnamefont{and}
  \bibinfo{author}{\bibfnamefont{P.}~\bibnamefont{Gegenwart}},
  \bibinfo{journal}{Phys. Rev. Lett.} \textbf{\bibinfo{volume}{110}},
  \bibinfo{pages}{176402} (\bibinfo{year}{2013}).

\bibitem[{\citenamefont{Millis et~al.}(2002)\citenamefont{Millis, Schofield,
  Lonzarich, and Grigera}}]{millis:prl-02}
\bibinfo{author}{\bibfnamefont{A.~J.} \bibnamefont{Millis}},
  \bibinfo{author}{\bibfnamefont{A.~J.} \bibnamefont{Schofield}},
  \bibinfo{author}{\bibfnamefont{G.~G.} \bibnamefont{Lonzarich}},
  \bibnamefont{and} \bibinfo{author}{\bibfnamefont{S.~A.}
  \bibnamefont{Grigera}}, \bibinfo{journal}{Phys. Rev. Lett.}
  \textbf{\bibinfo{volume}{88}}, \bibinfo{pages}{217204}
  (\bibinfo{year}{2002}).

\bibitem[{\citenamefont{Weickert et~al.}(2010)\citenamefont{Weickert, Brando,
  Steglich, Gegenwart, and Garst}}]{Weickert-prb10}
\bibinfo{author}{\bibfnamefont{F.}~\bibnamefont{Weickert}},
  \bibinfo{author}{\bibfnamefont{M.}~\bibnamefont{Brando}},
  \bibinfo{author}{\bibfnamefont{F.}~\bibnamefont{Steglich}},
  \bibinfo{author}{\bibfnamefont{P.}~\bibnamefont{Gegenwart}},
  \bibnamefont{and} \bibinfo{author}{\bibfnamefont{M.}~\bibnamefont{Garst}},
  \bibinfo{journal}{Phys. Rev. B} \textbf{\bibinfo{volume}{81}},
  \bibinfo{pages}{134438} (\bibinfo{year}{2010}).

\bibitem[{\citenamefont{Zacharias and Garst}(2013)}]{Zacharias:2013}
\bibinfo{author}{\bibfnamefont{M.}~\bibnamefont{Zacharias}} \bibnamefont{and}
  \bibinfo{author}{\bibfnamefont{M.}~\bibnamefont{Garst}},
  \bibinfo{journal}{Phys. Rev. B} \textbf{\bibinfo{volume}{87}},
  \bibinfo{pages}{075119} (\bibinfo{year}{2013}).

\bibitem[{\citenamefont{Fisher and Nelson}(1974)}]{Fisher:1974}
\bibinfo{author}{\bibfnamefont{M.}~\bibnamefont{Fisher}} \bibnamefont{and}
  \bibinfo{author}{\bibfnamefont{D.}~\bibnamefont{Nelson}},
  \bibinfo{journal}{Phys. Rev. Lett.} \textbf{\bibinfo{volume}{32}},
  \bibinfo{pages}{1350} (\bibinfo{year}{1974}).

\bibitem[{\citenamefont{Zhu et~al.}(2003)\citenamefont{Zhu, Garst, Rosch, and
  Si}}]{zhu}
\bibinfo{author}{\bibfnamefont{L.}~\bibnamefont{Zhu}},
  \bibinfo{author}{\bibfnamefont{M.}~\bibnamefont{Garst}},
  \bibinfo{author}{\bibfnamefont{A.}~\bibnamefont{Rosch}}, \bibnamefont{and}
  \bibinfo{author}{\bibfnamefont{Q.}~\bibnamefont{Si}}, \bibinfo{journal}{Phys.
  Rev. Lett.} \textbf{\bibinfo{volume}{91}}, \bibinfo{pages}{066404}
  (\bibinfo{year}{2003}).

\bibitem[{\citenamefont{Garst and Rosch}(2005)}]{GarstM:SigctG}
\bibinfo{author}{\bibfnamefont{M.}~\bibnamefont{Garst}} \bibnamefont{and}
  \bibinfo{author}{\bibfnamefont{A.}~\bibnamefont{Rosch}},
  \bibinfo{journal}{Phys. Rev. B} \textbf{\bibinfo{volume}{72}},
  \bibinfo{pages}{205129 } (\bibinfo{year}{2005}).

\bibitem[{\citenamefont{Tokiwa et~al.}(2009)\citenamefont{Tokiwa, Radu, Geibel,
  Steglich, and Gegenwart}}]{tokiwa-prl09}
\bibinfo{author}{\bibfnamefont{Y.}~\bibnamefont{Tokiwa}},
  \bibinfo{author}{\bibfnamefont{T.}~\bibnamefont{Radu}},
  \bibinfo{author}{\bibfnamefont{C.}~\bibnamefont{Geibel}},
  \bibinfo{author}{\bibfnamefont{F.}~\bibnamefont{Steglich}}, \bibnamefont{and}
  \bibinfo{author}{\bibfnamefont{P.}~\bibnamefont{Gegenwart}},
  \bibinfo{journal}{Phys. Rev. Lett.} \textbf{\bibinfo{volume}{102}},
  \bibinfo{eid}{066401} (\bibinfo{year}{2009}).

\bibitem[{\citenamefont{Tokiwa and Gegenwart}(2011)}]{tokiwa-rsi11}
\bibinfo{author}{\bibfnamefont{Y.}~\bibnamefont{Tokiwa}} \bibnamefont{and}
  \bibinfo{author}{\bibfnamefont{P.}~\bibnamefont{Gegenwart}},
  \bibinfo{journal}{Rev. Sci. Inst.} \textbf{\bibinfo{volume}{82}},
  \bibinfo{pages}{013905} (\bibinfo{year}{2011}).

\bibitem[{\citenamefont{Baier et~al.}(2007)\citenamefont{Baier, Steffens,
  Schumann, Kriener, Stark, Hartmann, Friedt, Revcolevschi, Radaelli, Nakatsuji, Maeno, Mydosh, Lorenz, and Braden}}]{Baier:2007}
\bibinfo{author}{\bibfnamefont{J.}~\bibnamefont{Baier}},
  \bibinfo{author}{\bibfnamefont{P.}~\bibnamefont{Steffens}},
  \bibinfo{author}{\bibfnamefont{O.}~\bibnamefont{Schumann}},
  \bibinfo{author}{\bibfnamefont{M.}~\bibnamefont{Kriener}},
  \bibinfo{author}{\bibfnamefont{S.}~\bibnamefont{Stark}},
  \bibinfo{author}{\bibfnamefont{H.}~\bibnamefont{Hartmann}},
  \bibinfo{author}{\bibfnamefont{O.}~\bibnamefont{Friedt}},
  \bibinfo{author}{\bibfnamefont{A.}~\bibnamefont{Revcolevschi}},
  \bibinfo{author}{\bibfnamefont{P.~G.} \bibnamefont{Radaelli}},
  \bibinfo{author}{\bibfnamefont{S.}~\bibnamefont{Nakatsuji}},
	\bibinfo{author}{\bibfnamefont{Y.}~\bibnamefont{Maeno}},
	\bibinfo{author}{\bibfnamefont{J. A.}~\bibnamefont{Mydosh}},
	\bibinfo{author}{\bibfnamefont{T.}~\bibnamefont{Lorenz}}, \bibnamefont{and}
	\bibinfo{author}{\bibfnamefont{M.}~\bibnamefont{Braden}},
  \bibinfo{journal}{J. Low Temp. Phys.}
  \textbf{\bibinfo{volume}{147}}, \bibinfo{pages}{405} (\bibinfo{year}{2007}).

\bibitem[{\citenamefont{Zacharias et~al.}(2009)\citenamefont{Zacharias,
  Woelfle, and Garst}}]{Zacharias:2009}
\bibinfo{author}{\bibfnamefont{M.}~\bibnamefont{Zacharias}},
  \bibinfo{author}{\bibfnamefont{P.}~\bibnamefont{Woelfle}}, \bibnamefont{and}
  \bibinfo{author}{\bibfnamefont{M.}~\bibnamefont{Garst}},
  \bibinfo{journal}{Phys. Rev. B} \textbf{\bibinfo{volume}{80}},
  \bibinfo{pages}{165116} (\bibinfo{year}{2009}).

\bibitem[{\citenamefont{Meng et~al.}(2012)\citenamefont{Meng, Rosch, and
  Garst}}]{Meng:2012}
\bibinfo{author}{\bibfnamefont{T.}~\bibnamefont{Meng}},
  \bibinfo{author}{\bibfnamefont{A.}~\bibnamefont{Rosch}}, \bibnamefont{and}
  \bibinfo{author}{\bibfnamefont{M.}~\bibnamefont{Garst}},
  \bibinfo{journal}{Phys. Rev. B} \textbf{\bibinfo{volume}{86}},
  \bibinfo{pages}{125107} (\bibinfo{year}{2012}).

\bibitem[{mar()}]{markus_note}
\bibinfo{note}{A vanishing of $\Gamma_{\rm H} \propto |h|^x$ for small $h$ with
  a non-integer exponent $x$ would imply a non-analytic behavior of the
  critical free energy as a function of small $h$: $\mathcal{F}_{\rm cr}
  \propto |h|^{x+1}$ with a $T$-dependent proportionality factor. This would
  give rise to a singular thermodynamics at $h=0$ for all $T$ because the
  $n^{\rm th}$-derivative $\partial^{(n)}_h \mathcal{F}_{\rm cr}$ with $x+1-n <
  0$ would then diverge for $|h| \to 0$. A singular line, i.e. a phase
  transition at $h=0$ is however not supported by experiment.}

\bibitem[{\citenamefont{Matsumoto et~al.}(2011)\citenamefont{Matsumoto,
  Nakatsuji, Kuga, Karaki, Horie, Shimura, Sakakibara, Nevidomskyy, and
  Coleman}}]{Matsumoto:2011}
\bibinfo{author}{\bibfnamefont{Y.}~\bibnamefont{Matsumoto}},
  \bibinfo{author}{\bibfnamefont{S.}~\bibnamefont{Nakatsuji}},
  \bibinfo{author}{\bibfnamefont{K.}~\bibnamefont{Kuga}},
  \bibinfo{author}{\bibfnamefont{Y.}~\bibnamefont{Karaki}},
  \bibinfo{author}{\bibfnamefont{N.}~\bibnamefont{Horie}},
  \bibinfo{author}{\bibfnamefont{Y.}~\bibnamefont{Shimura}},
  \bibinfo{author}{\bibfnamefont{T.}~\bibnamefont{Sakakibara}},
  \bibinfo{author}{\bibfnamefont{A.~H.} \bibnamefont{Nevidomskyy}},
  \bibnamefont{and} \bibinfo{author}{\bibfnamefont{P.}~\bibnamefont{Coleman}},
  \bibinfo{journal}{Science} \textbf{\bibinfo{volume}{331}},
  \bibinfo{pages}{316} (\bibinfo{year}{2011}).

\bibitem[{\citenamefont{Ramires et~al.}(2012)\citenamefont{Ramires, Coleman,
  Nevidomskyy, and Tsvelik}}]{Ramires:2012}
\bibinfo{author}{\bibfnamefont{A.}~\bibnamefont{Ramires}},
  \bibinfo{author}{\bibfnamefont{P.}~\bibnamefont{Coleman}},
  \bibinfo{author}{\bibfnamefont{A.~H.} \bibnamefont{Nevidomskyy}},
  \bibnamefont{and} \bibinfo{author}{\bibfnamefont{A.}~\bibnamefont{Tsvelik}},
  \bibinfo{journal}{Phys. Rev. Lett.} \textbf{\bibinfo{volume}{109}},
  \bibinfo{pages}{176404} (\bibinfo{year}{2012}).

\bibitem[{SM()}]{SM}
\bibinfo{note}{See supplementary material on the analysis of previously
  published data of thermal expansion \cite{Schmiedeshoff-prb11} and
  magnetization \cite{tokiwa-prb06}.}

\end{thebibliography}
\end{document}